# Coherent transfer of light polarization to electron spins in a semiconductor

Hideo Kosaka,[1,2] Hideki Shigyou,[1] Yasuyoshi Mitsumori,[1,2] Yoshiaki Rikitake,[2,3]
Hiroshi Imamura,[3,2] Takeshi Kutsuwa,[2] Koichiro Arai,[2,4] and Keiichi Edamatsu[1]

[1]*Laboratory for Nanoelectronics and Spintronics, Research Institute of Electrical Communication, Tohoku University, Sendai 980-8577, Japan*
[2]*CREST-JST, Saitama 332-0012, Japan*
[3]*Nanotechnology Research Institute, AIST, Tsukuba 305-8568, Japan*
[4]*ERATO Semiconductor Spintronics Project, JST, Saitama 332-0012, Japan*

We demonstrate that the superposition of light polarization states is coherently transferred to electron spins in a semiconductor quantum well. By using time-resolved Kerr rotation we observe the initial phase of Larmor precession of electron spins whose coherence is transferred from light. To break the electron-hole spin entanglement, we utilized the big discrepancy between the transverse g-factors of electrons and light holes. The result encourages us to make a quantum media converter between flying photon qubits and stationary electron spin qubits in semiconductors.

None of the physical implementations of quantum bits or qubits is suitable for all aspects of quantum information technology. Quantum information interfaces [1,2] connecting those qubits are thus needed to place the right qubit at the right position. Photons are the natural messenger qubits for fast and reliable communication [3], whereas electron spins in a semiconductor are suitable processor qubits for stable and scalable computation [4–6]. Quantum media conversion (QMC) between the photon qubit and the electron spin qubit will drastically extend the potential of quantum information and communication technologies.

Quantum coherence is one of the essential features of quantum phenomena and is a central issue in quantum information technology. Researchers exploring the feasibility of semiconductor-based quantum computation have demonstrated that not only the optical dipole coherence of excitons [7,8] but also electron spin coherence can be controlled optically [9–13] and electrically [5,6]. The electron spin coherence time, or transverse spin-relaxation time $T_2^*$ (the decay time of the quantum spin coherence), is generally much longer than the optical dipole coherence time and could be further prolonged by storing the electron spin coherence in a nuclear spin [14]. On the other hand, researchers exploring the feasibility of semiconductor-based quantum communication have shown that triggered single photons [15,16] and polarization-entangled photon pairs [17–20] can also be generated in semiconductors. One of the remaining challenges in developing all-semiconductor quantum info-communication devices is to transfer the quantum coherence of the superposition state of a photon (based on polarization degree of freedom) to that of an electron in a semiconductor (based on spin degree of freedom). Although the physics of optical orientation has a long history [21], little attention has been paid to the coherent transfer of the polarization state of light to electron spins.

In this letter we demonstrate for the first time coherent transfer of light polarization to electron spins. We not only show the projective up/down spin states or spin population as reported before [22–24] but also show that in a V-shaped three-level system with Voigt geometry an arbitrary coherent superposition state can be transferred from light polarization to electron spins. Although we used classical light and a spin ensemble, this coherent transfer is the first demonstration of the conditions needed for transferring the quantum state [25–27] of a photon to an electron.

We begin by summarizing the operating principle of our QMC scheme (Fig. 1). Any light polarization state is represented by a state vector, called a Stokes vector, in the Poincaré sphere (PS) spanned by the basis vectors $|\sigma^+\rangle$ and $|\sigma^-\rangle$, corresponding to right and left circular polarizations. In the same manner, any electron spin state is represented by a Bloch vector in the Bloch sphere (BS) spanned by the basis vectors $|\uparrow\rangle_e$ and $|\downarrow\rangle_e$, corresponding to up and down spin polarizations. Both the PS and BS thus are equivalent SU(2) Hilbert spaces connected via selection rules. The selection rules for a light hole (LH) exciton are the following [1]:

$$|\sigma^+\rangle \rightarrow |\uparrow\rangle_e \otimes |\downarrow\rangle_{lh}, \quad |\sigma^-\rangle \rightarrow |\downarrow\rangle_e \otimes |\uparrow\rangle_{lh}. \quad (1)$$

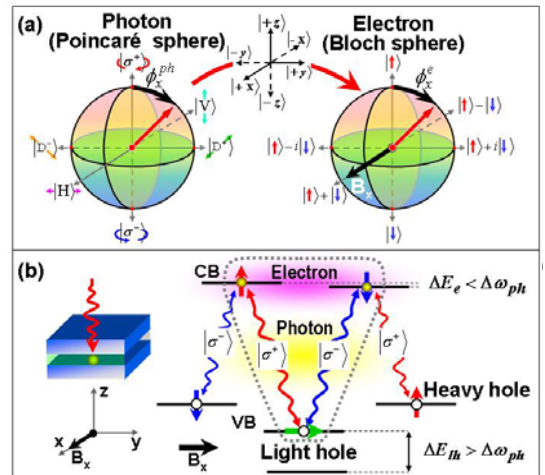

FIG. 1. The operating principle of our QMC scheme from photon polarization to electron spin polarization. (a) State vectors representing a photon polarization in a Poincaré sphere and an electron spin polarization in a Bloch sphere with the phase definitions used in Fig. 3. (b) Schematic diagrams of the experimental setup and the relevant energy levels together with optical spin selection rules under an in-plane magnetic field $B_x$ that forms a V-shaped three-level system consisting of a degenerate electron state and one eigenstate of non-degenerate light-hole (LH) states. The photon bandwidth ($\Delta\omega_{ph}$) should be larger than the electron Zeeman splitting ($\Delta E_e$) but smaller than the LH Zeeman splitting ($\Delta E_{LH}$).

An arbitrary light polarization state is thus transferred to the exciton spin state as

$$\alpha|\sigma^+\rangle + \beta|\sigma^-\rangle \rightarrow \alpha|\uparrow\rangle_e \otimes |\downarrow\rangle_{lh} - \beta|\downarrow\rangle_e \otimes |\uparrow\rangle_{lh}. \quad (2)$$

One can see in Eq. (2) that the electron spin state is entangled with the hole spin state. The coherence time of a hole spin, however, is so short that the entangled electron spin might be instantly converted into incoherent mixture of spin states projected along the z axis, and an incoherent mixture cannot be used as a qubit. The problem can be solved by applying an in-plane (transverse) magnetic field $B_x$ to configure the V-shaped three-level system shown in Fig. 1(b). The magnetic field lifts the Kramers degeneracy of the LH and reconfigures the coupled eigenstates as $|\pm x\rangle_{lh} = |\downarrow\rangle_{lh} \pm |\uparrow\rangle_{lh}$ while keeping the degeneracy of the electron with the smaller g-factor. Via the resonant transition between the $|-x\rangle_{lh}$ state and the degenerated electron states, the same selection rules lead to the following equation:

$$\alpha|\sigma^+\rangle + \beta|\sigma^-\rangle \rightarrow (\alpha|\uparrow\rangle_e + \beta|\downarrow\rangle_e) \otimes |-x\rangle_{lh}. \quad (3)$$

Since the transferred electron spin state in Eq. (3) is not entangled with the LH state, its coherence time is not limited by the short LH spin coherence time or the fast relaxation to the ground heavy hole (HH) state. This is the key point of our QMC scheme. It is essential to use the LH states since we cannot sufficiently lift the Kramer's degeneracy of the HH state by applying an in-plane magnetic field [28]. The electron coherence time is limited by the electron-hole exchange interaction, which can be greatly extended by extracting a hole while leaving an electron in a quantum well [9] or a quantum dot [23,29].

The sample we used contained a high-quality 11-nm-thick undoped GaAs quantum well embedded in undoped $Al_{0.3}Ga_{0.7}As$ grown by molecular beam epitaxy on a semi-insulating GaAs substrate with the growth interruption to have a high quality well. The photoluminescence (PL) spectrum of the HH exciton shown in Fig. 2(a) has three structures that correspond to the different confinement energies caused by the one-monolayer fluctuation of the well width. The homogeneous linewidth should be much narrower than the inhomogeneous linewidth of 1 nm (2 meV). The photoluminescence excitation (PLE) spectrum in Fig. 2(a) shows that the LH exciton is energetically well separated from the HH exciton. The quantum well was designed to have a relatively small electron transverse g-factor ($g_{e,\perp} = -0.21$) and a relatively large LH transverse g-factor ($g_{lh,\perp} = -3.5$) under the in-plane magnetic field $B_x$ used to configure the V-shaped three-level system. We chose the non-zero g-factor for an electron intentionally in order to be able to observe the initial phase of the spin precession under a moderate $B_x$. A mode-locked Ti:Sapphire laser (Coherent MIRA 900) delivering 130-fs pulses at a repetition rate of 76 MHz was used to pump and probe the sample through a wavelength-tunable filter consisting of a diffraction grating (FWHM = 0.38 nm) in each path and a variable delay line in only the probe path. All the measurements were made at 5 K and at normal incidence in the Voigt geometry (Fig. 1(b)). The electron spin dynamics free from the influence of the dynamic nuclear-spin polarization [13,14] was extracted by periodically switching the polarization of the pump light between two orthogonal states. The pump power dependence of the typical Kerr rotation signals (Fig. 2(b)) is linear up to 1 mW, which corresponds to the exciton density of $5 \times 10^9$ cm$^{-2}$ or an average exciton spacing of 160 nm, which is much greater than the exciton Bohr radius (< 12 nm). Below this power level, all the temporal responses were unchanged. All the measurements in the work reported here were made in this linear range where created excitons were well separated from each other.

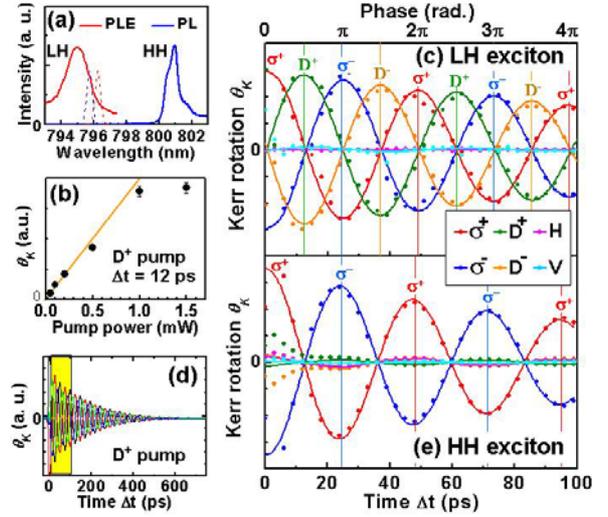

FIG. 2. (a) PL (blue) and PLE (red) spectra detected at HH exciton emission under $B_x = 7$ T. The broken lines show the spectra of the pump (red) and probe (blue) lights. (b) Pump power dependence of the Kerr rotation angle ($\theta_K$) with $D^+$ under $B_x = 7$ T at $\Delta t = 12$ ps. Error bars show s.d. (c) Temporal evolution of $\theta_K$ with six basis polarizations under $B_x = 7$ T. Each of the consecutive phase shifts in the cyclic order of $|\sigma^+\rangle$ (red)→$|D^+\rangle$ (green)→$|\sigma^-\rangle$ (blue)→$|D^-\rangle$ (orange)→$|\sigma^+\rangle$ in $\pi/2$ steps coincides well with that of electron spin states expected from the pump-light polarizations. In contrast, excitation with the $|H\rangle$ (pink) and $|V\rangle$ (sky blue) polarization, which correspond to the $B_x$ direction, generate negligible signals. (d) Large-time-scale view of Fig. 2(c). (e) HH-excitation case for comparison. Only $|\sigma^\pm\rangle$ excitations generate relatively large signals, indicating that spins are always projected along the z axis.

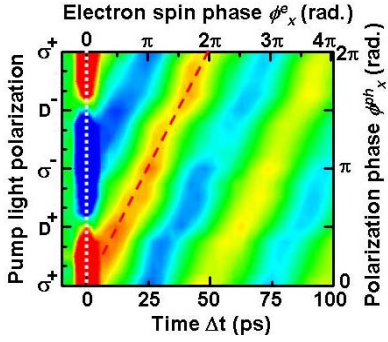

FIG. 3. Coincidence of the state vector angle between the light polarization and the electron spin polarization. Time evolution of $\theta_K$ is shown as a function of pump light polarization under $B_x = 7$ T. The linear streak pattern indicates coincidence of the phase angle of the electron spins $\phi_x^e$ to that of the light polarization $\phi_x^{ph}$. Definitions of the phases are shown in Fig. 1(a).

Quantum spin coherence of an electron is often indicated by the Larmor precession or the quantum beats [9-12,22,28] originating from the quantum interference of two eigenstates, which is usually visualized by time-resolved Kerr rotation measurements [9,10]. Although the Kerr rotation provides information only about the spin projection or population along the growth axis, we can infer the spin orientation of the superposition state from the initial phase of the spin precession. Here we show how the inferred initial phase of the generated electron spin is related to the pump light polarization. We first selected six light polarization states $\{|\sigma^+/\sigma^-\rangle, |D^+/D^-\rangle\ (\pm 45°\ \text{degree polarization})$ and $|H/V\rangle$ (horizontal/vertical polarization)} and measured the temporal evolution of the Kerr rotation angle $\theta_K$ under $B_x = 7$ T (Fig. 2(c)). The lower-energy side of the LH exciton peak ($\lambda_E = 796.2$ nm) was excited with one of the above-mentioned polarizations and probed at $\lambda_P = 795.7$ nm with the H polarization (Fig. 2(a)). Four of the states perpendicular to the x axis ($|\sigma^+\rangle, |D^+\rangle, |\sigma^-\rangle$ and $|D^-\rangle$) provided approximately the same amplitude, while the phases were sequentially shifted by the angle of $\pi/2$. These results suggest that the $|D^\pm\rangle$ states, which are coherent superpositions of $|\sigma^+\rangle$ and $|\sigma^-\rangle$ states $|\sigma^+\rangle \pm i\ |\sigma^-\rangle$, create equivalent superpositions of the $|\uparrow\rangle_e$ and $|\downarrow\rangle_e$ states $|\uparrow\rangle_e \pm i\ |\downarrow\rangle_e$ (normalization omitted for brevity). It is natural that the states along the x axis ($|H\rangle$ and $|V\rangle$) provided negligible amplitude because those state vectors are parallel to the field. From the decay of precession amplitude seen in Fig. 2(d), the electron spin coherence time $T_2^*$ time is estimated to be 160 ps. In contrast to the LH excitation, the HH excitation ($\lambda_E = \lambda_P = 800.8$ nm) gave the conventional result (Fig. 2(e)), where the $\sigma^\pm$ pump corresponding to the projected states $|\uparrow/\downarrow\rangle_e$ provided significant amplitude but the others did not. The lack of the precession amplitude seen with $D^+$ and $D^-$ pumping indicates that superposition of the $|\uparrow\rangle_e$ and $|\downarrow\rangle_e$ states could not be created by the HH excitation.

The major advantage of quantum bits over classical bits is that they can take any superposition states of a two-level system. Arbitrary light polarization states therefore have to be coherently transferred to well-defined electron spin states that should show one-to-one correspondence. To show this correspondence, we verified dependence of the electron spin phase $\phi_x^e$ on the light polarization phase $\phi_x^{ph}$ as shown in Fig. 3. The phases $\phi_x^e$ and $\phi_x^{ph}$ are respectively the angles of the state vectors in the PS and the BS, such that $|D^+\rangle = \cos(\phi_x^{ph}/2)\ |\sigma^+\rangle + i\ \sin(\phi_x^{ph}/2)\ |\sigma^-\rangle\ (\phi_x^{ph} = \pi/2)$ as shown in Fig. 1(a). We can infer the phase $\phi_x^e$ from the peaks tracked by the dashed red line. The inferred electron phases $\phi_x^e$ well coincide with the pump-light phases $\phi_x^{ph}$. The coincidence of the orientation of the state vectors in both the PS and BS is evidence of coherent polarization transfer.

To show the role of the lift of Kramers degeneracy induced by the in-plane magnetic field, we measured $B_x$-field dependence of $\theta_K$ with the $D^+$ polarization (Fig. 4(a)). The times to reach maximum and minimum signals are well reproduced by the theoretical calculations shown by the dashed curves. The first maximum points are delayed by the phase angle of $\pi/2$ for all $B_x$ as in Fig. 2(c). Note that as the magnetic field decreases, the precession amplitude decreases much faster than the spin $T_2^*$ time, indicating that the coherence of polarization of electron spins is gradually lost through the entanglement with the hole state as is the case of the HH excitation. The precession amplitude is observed only when selectively exciting one eigenstate of the hole spin states split by an in-plane field. The distinguishability of the LH states, which is essential for the QMC, is demonstrated in Fig. 4(b), where the $\theta_K$ is measured at $\Delta t = 12$ ps (corresponding to the $\pi/2$ spin rotation) with the $D^+$ polarization at $B_x = 7$ T is plotted against both the pump and probe wavelengths. The splitting of the peaks along the pump wavelength originates from the LH Zeeman splitting. The low (high)-energy peak showing the positive (negative) sign corresponds to the LH spin state $|\pm x\rangle_{lh} = |\uparrow\rangle_{lh} \pm |\downarrow\rangle_{lh}$, where the pump-light polarization state $|D^+\rangle$ is transferred to the electron spin state $|\pm y\rangle_e = |\uparrow\rangle_e \pm i|\downarrow\rangle_e$, respectively. We can fit the pump spectrum of $\theta_K$ (dots) at a fixed probe wavelength by using two Gaussian functions with a separation given by the calculated transverse LH g-factor of $-3.5$ (solid curve). This well-separated spectrum is not observed in the HH excitation case, where because of the Kramer's degeneracy only one peak is observed with any polarization of pump light.

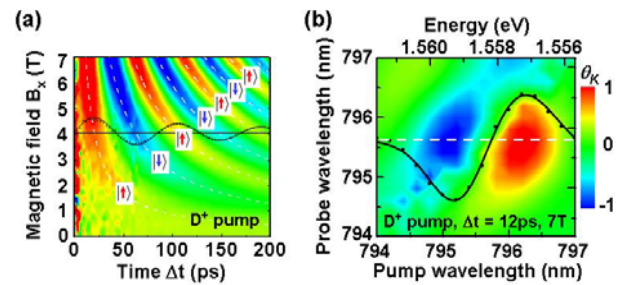

FIG. 4. (a) Time evolution of $\theta_K$ as a function of $B_x$ with $D^+$ pumping. The cut of $\theta_K$ at $B_x = 4$ T is shown by the dots fitted to a damped sine function. The dashed curves are the calculated tracks of the $B_x - \Delta t$ values needed to detect electrons aligned to $|\pm z\rangle_e$ when an electron transverse g-factor of $-0.21$ is assumed. (b) $\theta_K$ as a function of pump/probe wavelengths with $D^+$ at $\Delta t = 12$ ps under $B_x = 7$ T. Dots show the pump spectrum of $\theta_K$ at the probe wavelength used for all the LH measurements (795.7 nm) with a curve fitted by a two-component Gaussian.

We also confirmed that none of the above experimental results were changed by field reversal. The physical interpretation of this constancy is that field reversal acts as a time reversal together with a space inversion, leading to the same observation. For example, when pumped with the $D^+$ polarization, the generated electron spin state $|+y\rangle_e$ flips to $|-y\rangle_e$ when the field changes from $+B_x$ to $-B_x$ because the excited hole state changes from $|-x\rangle_{lh}$ to $|+x\rangle_{lh}$. At the same time, the g-factor behaves as if the sign is flipped, leading to the reversed spin rotation. Consequently, from the spin projection along the z axis we cannot distinguish these two cases.

The coherent polarization transfer demonstrated here is essentially different from the previously demonstrated coherent electron-spin controls based on an optical Stark effect [10], spin-flip Raman scattering [11] or Rabi oscillation [12], each of which needs intense light as a driving force defined and is therefore not suitable for transferring the phase state of a single photon. Our QMC scheme is not limited to the transfer of a single-particle state but can be generalized to the transfer of a two-particle state or an entangled state, which is the kind of transfer needed for quantum repeaters [30,31] or routers for scalable quantum communication and distributed quantum computation. Although the wavelength used here is about half of the telecom wavelength, QMC between photons of different wavelengths by a nonlinear up-conversion process has been shown to be possible [2]. The transfer of light polarization only in the y-z plane in the PS as shown here is sufficient for implementation of a quantum repeater based on the BB84 cryptographic scheme [2,3] by converting the time-bin phase coding to the polarization coding.

In conclusion, we have demonstrated coherent polarization transfer from light to electron spins in a semiconductor. From the phase of spin precession we inferred the initial phase that should correspond to the phase of light polarization. The electron-hole spin entanglement was broken by the big discrepancy between the g-factor of the electron and the g-factor of the light hole. Although we have demonstrated only a necessary condition of QMC, this is a step toward the realization of all-semiconductor quantum interfaces between a photon and an electron spin.

We are grateful to Professors Hideo Ohno, Toshihide Takagahara, Keiji Ono and Jaw-Shen Tsai for fruitful discussions. This work was supported in part by the Strategic Information and Communications R & D Promotion Program (SCOPE No. 41402001) of the Ministry of Internal Affairs and Communications, a Grant-in-Aid for Scientific Research (No. 16710061) from the Ministry of Education, Culture, Sports, Science and Technology (MEXT), and a grant from the International Joint Research Program of the New Energy and Industrial Technology Development Organization (NEDO).

**REFERENCES**

[1] R. Vrijen and E. Yablonovitch, *Physica E* **10**, 569 (2001).
[2] S. Tanzilli et al., *Nature* **437**, 116 (2005).
[3] C. H. Bennett and G. Brassard, in *Proceedings of the International Conference on Computers, Systems and Signal Processing* (IEEE, New York, 1984), p. 175.
[4] D. Loss and D. P. DiVincenzo, *Phys. Rev. A* **57**, 120 (1998).
[5] J. R. Petta et al., *Science* **309**, 2180 (2005).
[6] F. H. L. Koppens et al., *Nature* **442**, 766 (2006).
[7] N. H. Bonadeo et al., *Science* **282**, 1473 (1998).
[8] A. Zrenner et al., *Nature* **418**, 612 (2002).
[9] J. M. Kikkawa and D. D. Awschalom, *Nature* **397**, 139 (1999).
[10] J. A. Gupta, R. Knobel, N. Samarth and D. D. Awschalom, *Science* **292**, 2458 (2001).
[11] M. V. G. Dutt et al., *Phys. Rev. Lett.* **94**, 227403 (2005).
[12] A. Greilich et al., *Phys. Rev. Lett.* **96**, 227401 (2006).
[13] M. Atatüre et al., Science **312**, 551 (2006).
[14] M. V. G. Dutt et al., Science **316**, 1312 (2007).
[15] C. Santori, M. Pelton, G. Solomon, Y. Dale and Y. Yamamoto, *Phys. Rev. Lett.* **86**, 1502 (2001).
[16] Z. Yuan et al., *Science* **295**, 102 (2002).
[17] K. Edamatsu, G. Oohata, R. Shimizu and T. Itoh, *Nature* **431**, 167 (2004).
[18] R. M. Stevenson et al., *Nature* **439**, 179 (2006).
[19] N. Akopian et al., *Phys. Rev. Lett.* **96**, 130501 (2006).
[20] G. Oohata, R. Shimizu and K. Edamatsu, *Phys. Rev. Lett.* **98**, 140503 (2007).
[21] F. Meier and B. P. Zakharchenya, Eds., *Optical Orientation* (Elsevier, Amsterdam, 1984).
[22] A. P. Heberle, W. W. Ruhle and K. Ploog, *Phys. Rev. Lett.* **72**, 3887 (1994).
[23] M. Kroutvar et al., *Nature* **432**, 81 (2004).
[24] H. Kosaka, Y. Mitsumori, Y. Rikitake and H. Imamura, *Appl. Phys. Lett.* **90**, 113511 (2007).
[25] J. I. Cirac, P. Zoller, H. J. Kimble and H. Mabuchi, *Phys. Rev. Lett.* **78**, 3221 (1997).
[26] D. N. Matsukevich and A. Kuzmich, *Science* **306**, 663 (2004).
[27] J. F. Sherson et al., *Nature* **443**, 557 (2006).
[28] X. Marie et al., *Phys. Rev. B* **60**, 5811 (1999).
[29] H. Kosaka et al., *Phys. Rev. B* **67**, 045104 (2003).
[30] H.-J. Briegel, W. Dür, J. I. Cirac and P. Zoller, *Phys. Rev. Lett.* **81**, 5932 (1998).
[31] P. van Loock et al., *Phys. Rev. Lett.* **96**, 240501 (2006).